\documentclass[aip,jcp,preprint,floatfix]{revtex4-1}

\usepackage{hyperref}
\usepackage{graphicx}
\usepackage{amstext}
\usepackage{amsmath}
\usepackage{amssymb} 
\usepackage{color}
\usepackage{changes}
\usepackage{float}

\def\be{\begin{equation}}
\def\ee{\end{equation}}
\def\bea{\begin{eqnarray}}
\def\eea{\end{eqnarray}}
\def\bal#1\eal{\begin{align*}#1\end{align*}}
\def\ba#1\ea{\begin{align}#1\end{align}}

\newcommand{\bra}[1]{\langle #1|}
\newcommand{\ket}[1]{| #1\rangle}
\newcommand{\fig}[1]{Fig.~\ref{#1}}

\newcommand{\eq}[1]{Eq.~\eqref{#1}}

\begin{document}
\title{Illustrating Electric Conductivity Using the Particle-in-a-Box Model: Quantum Superposition is the Key }

\author{Umaseh Sivanesan}
\affiliation{Department of Physical and Environmental Sciences,
  University of Toronto Scarborough, Toronto, Ontario, M1C 1A4,
  Canada}
\author{Kin Tsang}
\affiliation{Department of Physical and Environmental Sciences,
  University of Toronto Scarborough, Toronto, Ontario, M1C 1A4,
  Canada}
\author{Artur F. Izmaylov} %
\email{artur.izmaylov@utoronto.ca}
\affiliation{Department of Physical and Environmental Sciences,
  University of Toronto Scarborough, Toronto, Ontario, M1C 1A4,
  Canada; and Chemical Physics Theory Group, Department of Chemistry,
  University of Toronto, Toronto, Ontario, M5S 3H6, Canada}

\begin{abstract}
Most of the textbooks explaining electric conductivity in the context of quantum mechanics 
provide either incomplete or semi-classical explanations that are not 
connected with the elementary concepts of quantum mechanics.  
We illustrate the conduction phenomena using the simplest model system in quantum dynamics, a particle in a box (PIB). 
To induce the particle dynamics, a linear potential tilting the bottom of the box is introduced,
which is equivalent to imposing a constant electric field for a charged particle.
Although the PIB model represents a closed system that cannot have a flow of electrons 
through the system, we consider the oscillatory 
dynamics of the particle probability density as the analogue of the electric current. 
Relating the amplitude and other parameters of the particle oscillatory dynamics 
with the gap between the ground and excited states of the PIB model allows us  
to demonstrate one of the most basic dependencies of electric conductivity on 
the valence-conduction band gap of the material.
\end{abstract}
\maketitle

\section{Introduction}
Electric conductivity is an inherently quantum phenomenon because it depends on the
energy level structure present in materials. Although electron orbital energies in 
materials form continuous bands, the energy gap between occupied orbitals (valence band)
and unoccupied orbitals (conduction band) can be finite. 
This gap predominantly defines the conductive properties of any material: 
the band gap of conductive materials is small (semi-conductors) or 
zero (metals), while nonconductive materials (insulators) have a large gap. 

Electric conductivity is a dynamical phenomenon and in quantum mechanics 
it must be represented by the dynamics of a non-stationary state. Thus if one would like to 
provide a simple, idealized, zero-temperature illustration within quantum mechanics 
without going into quantum statistical thermodynamics, the system needs to be in a superposition state. 
Unfortunately, in the desire for simplicity, 
most of the textbooks go too far passing by the concept of the superposition state
and produce awkward explanations, where under the electric field potential,
an electron can move and at the same time be in a stationary state.\cite{Fayer:2013/73,Engel:2013/73}
This is of course a misleading oversimplification, because the probability density for any object 
in a stationary state is time-independent.  

In order to provide a simple but quantum mechanically correct 
explanation for the dependence of conductance on the band gap, 
one needs to consider a superposition state at some point. Moreover, since the superposition 
is the key to electronic motion under the influence of electric field, its creation 
should be more pronounced when the gap between ground and excited levels 
is smaller. On the other hand increasing the gap should reduce the potency of 
the electric field to create a superposition state.

Some solid-state textbooks\cite{Ziman:1960/92,Ashcroft:1976/214} provide an 
explanation that involves electronic superposition states 
or wave packets built as a linear combination of one-electron Bloch functions 
obtained for periodic potentials. Of course, such explanations would not be possible to 
incorporate in undergraduate quantum mechanics courses without significant 
detours into the consideration of periodic systems.

In this paper we present a simple single-particle illustration of the electric conduction 
phenomena based on the one-dimensional particle-in-a-box (PIB) model. 
The key element of our consideration is the superposition state emerging 
under the influence of an electric field, modelled as a sudden tilt of the 
potential box's bottom. This superposition state gives rise to the quantum dynamics 
necessary for particle transport within the box. Properties of this superposition 
state are related to the gap between occupied and unoccupied 
energy levels, thus providing a simple explanation of the gap-conduction relation in real materials.  
There are a few simplifications separating dynamics of our model system from the conductance dynamics 
in the real material: 1) only one electron is considered and thus electron-electron and electron-nuclear interactions 
are neglected; 2) the PIB model represents a closed system, and its dynamics cannot produce steady 
particle current; instead, particle's movement has an oscillatory character. 
These simplifications are not essential for illustrating the gap-conductance relation
if we associate the conductance with various dynamical properties of a single-particle probability density. 

This view of conduction can be taught at the elementary level of quantum mechanics 
for undergraduate students. All illustrations provided in this paper
can be demonstrated in the class or given as separate projects for 
advanced undergraduate or even graduate courses. 

The rest of the paper is organized as follows. In Sec. II.A we consider the PIB model and the introduction of a 
constant electric field. Section II.B-D describe various properties characterizing dynamics in the model using 
perturbative considerations. We illustrate all discussions in Sec. III by simulating the dynamical quantities employing the
variational method for the PIB model with an inclined bottom.  
For simplicity of involved expressions, atomic units are used throughout this paper.

\section{Theory}

\subsection{Particle in a box model}

To establish notation let us introduce a particle of unit charge with mass $m$ in a box of size $L$ 
with infinite potential walls, then the Hamiltonian is 
\bea
H_0(x) &=& -\frac{1}{2m}\frac{d^2}{dx^2}+V_0(x),\\
V_0(x) &=& \left\{ \begin{array}{ll} 0,~ x \in[0,L] \\ 
\infty,~x\not\in[0,L].\end{array} \right.
\eea
$H_0$ defines the time-independent Schr\"odinger equation for the stationary states
\ba
H_0(x)\psi_n^{(0)}(x)=E_n^{(0)}\psi_n^{(0)}(x),
\ea
where the eigenenergies and eigenfunctions are
\bea
E_n^{(0)} &=& \frac{n^2\pi^2}{2mL^2}, \\
\psi_n^{(0)}(x) &=& \sqrt{\frac{2}{L}}\sin\left({\frac{n{\pi}x}{L}}\right),
\eea
and they are enumerated by the subscript $n$, where $n$=1 corresponds to the ground state.

We will use the ground state as the initial state of our system in the absence of the electric field.
To model the electric field that creates bias, we add to PIB's Hamiltonian a linear potential within the box
\bea
V(x) &=& \left\{ \begin{array}{ll} -\mathcal{E}x,~ x \in[0,L] \\ 
\infty,~x\not\in[0,L],\end{array} \right.
\eea 
so that the total system Hamiltonian with the electric field becomes
\ba\label{eq:H}
H(x) = H_0(x) +V(x).
\ea
This linear potential can be seen as a result of applying a constant electric field $\mathcal{E}$ 
because the electric potential $\varphi(x)$ entering the Hamiltonian for a charged particle will be 
exactly $-\mathcal{E} x$.

Upon sudden turn-on of the electric field,  
our original state $\psi(x,t=0) = \psi_1^{(0)}(x)$ is not a stationary state of $H(x)$, it is instead a 
superposition of $H(x)$'s eigenstates and thus undergoes dynamics. 
The simplest way to obtain this dynamics is using the eigenstates of $H(x)$
\ba \label{eq:tise2}
H(x) {\psi}_n(x) = {E}_n {\psi}_n(x)
\ea
to solve the time-dependent Schr\"odinger equation (TDSE). 
Expanding the initial state 
\bea
\psi(x, 0) &=& \psi_1^{(0)}(x) \\ \label{eq:ini}
&=& \sum_{n=1}^{\infty}c_n{\psi}_n(x)
\eea
as a linear combination of ${\psi}_n(x)$ 
with coefficients 
\ba\label{eq:cn}
c_n = \int_{0}^{L}dx{\psi}^\ast_n(x)\psi_1^{(0)}(x)
\ea
allows us to write a solution of TDSE as
\ba\label{eq:wft}
\psi(x, t) = \sum_{n=1}^{\infty}c_n{\psi}_n(x)e^{-i{E}_{n}t}.
\ea
This wave function has a time-dependent probability distribution 
\ba\label{eq:dist}
|\psi(x,t)|^2 =  \sum_{n,k} c_n^* c_k {\psi}_n^{*}(x){\psi}_k(x)e^{i({E}_{n}-{E}_k) t}.
\ea
The key element of this time-dependence is the superposition character of the 
original wave function, thereby leading to time-dependent exponents in \eq{eq:dist}.
In contrast, if $\psi(x,0)$ consisted of a single eigenstate ${\psi}_n(x)$, its 
 time-dependence $\psi(x,t) = {\psi}_n(x)\exp(-i{E}_nt)$ would not be present in 
 the probability density $|\psi(x,t)|^2 = |{\psi}_n(x)|^2$.
 
In our model, no matter how small the electric field is, $\psi(x,0)=\psi_1^{(0)}(x)$ 
will always be a non-stationary function for the total Hamiltonian. Thus the dynamics
 will always be present; the only question is whether the changes in localization of 
 the probability distribution over time are significant. It is clear from \eq{eq:dist} that for significant 
 dynamics, the amplitudes of the cross terms $c_n^* c_k$ ($n\ne k$) must be large.
 
 To obtain $c_n$'s and $E_n$'s needed for simulation \eq{eq:dist} one can use the variational 
 approach, which searches for extrema of the energy functional with respect to variation of a
 trial wave function $\psi(x)$
\ba
{E} = \frac{\bra{\psi}H\ket{\psi}}{\bra{\psi}\psi\rangle}.
\ea
The simplest form of the trial wave function is a linear parameterization 
 \bea
 \psi(x) = \sum_{n=1}^{N} C_n \psi_n^{(0)}(x), 
 \eea
 where $C_n$'s can be varied to find extrema of $E$. 
 This variation leads to a system of $N$ equations obtained from the $\frac{\partial E}{\partial C_n} = 0$ 
 conditions which is equivalent to the eigenvalue problem 
\ba\label{eq:eig}
\mathbf{HC} = E\mathbf{C},
\ea
where $\mathbf{H}$ is the Hamiltonian matrix with the elements
\bea
H_{nk} &=& \int_0^{L} \psi_n^{(0)*}(x) \hat H \psi_k^{(0)}(x)dx \\
&=& \left\{ \begin{array}{ll}  \frac{4\mathcal{E}nkL[1-(-1)^{n+k}]}{\pi^2(n-k)^2(n+k)^2},~ n \neq k \\ 
\frac{n^2\pi^2}{2mL^2} - \frac{\mathcal{E}L}{2},~ n = k,\end{array} \right.
\eea
and $\mathbf{C} =(C_1,C_2,...C_N)^T$ is an eigenvector corresponding to an eigen-energy $E$.
There are $N$ eigenvectors  and $N$ eigenenergies in total; the $n$'th $E$ corresponds to an approximation 
to $E_n$ in \eq{eq:tise2}, and $C_1$ for the $n$'th $\mathbf{C}$ is equal to $c_n$ in \eq{eq:ini}.
Therefore, solving the eigenvalue problem in \eq{eq:eig} gives all needed quantities to construct the dynamics
of the initial distribution under the influence of the sudden turn-on of the constant electric field using \eq{eq:dist}. 

\subsection{Dynamics of probability density}
To obtain some qualitative insight in what determines the probability density dynamics, we will 
consider which system parameters affect $c_n$ coefficients, because the larger the spread of 
these coefficients, the more pronounced the expected dynamics.
We will use time-independent perturbation
theory (TIPT) as our main tool in this analysis. TIPT formulates eigenstates of $H(x)$ as a series
\ba
\psi_n(x) = \psi_n^{(0)}(x) +  \psi_n^{(1)}(x) + \psi_n^{(2)}(x)+ ...,
\ea
where $\psi_n^{(0)}(x)$ are eigenfunctions of the $H_0$ Hamiltonian and $\psi_n^{(k)}(x)$ 
are higher order perturbative corrections to $\psi_n^{(0)}(x)$.
If we substitute this expansion in \eq{eq:cn} we can obtain the corresponding expansion for $c_n$
\ba
c_n &= c_n^{(0)} +  c_n^{(1)} + c_n^{(2)} +... \\ \label{eq:cnk}
c_n^{(k)} &= \int_0^L dx  [\psi_n^{(k)}(x)]^{*}\psi_1^{(0)}(x).
\ea
For weak electric fields we can neglect all terms beyond the first order. Note that due to 
the zeroth-order functions' orthogonality, we only have the zeroth-order term for $n=1$.
For $n\ne 1$, the first nontrivial term is $c_n^{(1)}$. In the non-degenerate case 
corresponding to $H_0$, TIPT provides the following expression for 
$\psi_n^{(1)}(x)$,
\ba
\psi_n^{(1)}(x) &= \sum_{k\ne n} \frac{\bra{\psi_k^{(0)}}V\ket{\psi_n^{(0)}}}{E_n^{(0)}-E_k^{(0)}} \psi_k^{(0)}(x),
\ea  
Substituting this equation into \eq{eq:cnk} and using the orthogonality condition 
$\bra{\psi_k^{(0)}}\psi_1^{(0)}\rangle=\delta_{k1}$ gives
\ba
c_{n}^{(1)} = \frac{\bra{\psi_n^{(0)}}V\ket{\psi_1^{(0)}}}{E_1^{(0)}-E_n^{(0)}},~n\ne1.
\ea
This expression clearly shows that $c_n^{(1)}$ are inversely proportional to the gap between ground and $n$'th
state $E_1^{(0)}-E_n^{(0)}$. The numerator and denominator of this expression can be evaluated analytically for our model
\ba\label{eq:Vn1}
\bra{\psi_n^{(0)}}V\ket{\psi_1^{(0)}} &= \frac{4n\mathcal{E}L[(-1)^{n}+1]}{\pi^2(n^2-1)^2},
\\\label{eq:gap}
E_1^{(0)}-E_n^{(0)} &= \frac{\pi^2(1-n^2)}{2mL^2}.
\ea
Putting all the components together we obtain
\ba
c_{n}^{(1)} = \frac{-8mn\mathcal{E}L^3[(-1)^{n}+1]}{\pi^4(n^2-1)^3}{\label{eq:cn1}},~n\ne1.
\ea
To characterize a non-stationary character of the initial distribution, one can use 
the sum over weights of all excited states
\ba
\omega_{\rm exc} &= \sum_{n=2}^{\infty} |c_n|^2\approx \sum_{n=2}^{\infty} |c_n^{(1)}|^2\\
&= \sum_{n=2}^{\infty}\frac{64m^2\mathcal{E}^2n^2L^6[(-1)^{n}+1]^2}{\pi^8(n^2-1)^6}.
\ea
Here, all odd $n$'s are zero, thus only even terms ($n=2j$) need to be considered 
\ba
\omega_{\rm exc} &= \frac{1024m^2\mathcal{E}^2L^6}{\pi^8}\sum_{j=1}^{\infty}\frac{j^2}{(4j^2-1)^6}\\
\label{eq:omega}
&= \frac{(2\pi^4+5\pi^2-210)m^2\mathcal{E}^2L^6}{240\pi^6}.
\ea
The infinite sum has been determined by the {\it Mathematica} program.\cite{Mathematica}
Therefore, perturbation theory predicts growth of $\omega_{\rm exc}$ as $m^2\mathcal{E}^2L^6$.

\subsection{Polarizability}

An alternative way to characterize the mobility of the initial distribution is to calculate the polarizability of 
the system. The polarizability of the ground state is the second derivative of the energy with respect to the electric field
\ba{\label{eq:pol}}
\alpha = \frac{d^2E_1}{d\mathcal{E}^2} \Big{\vert}_{\mathcal{E}=0}.
\ea  
TIPT is exactly the right tool to obtain such derivatives.\cite{Kutzelnigg:2006ij} 
According to TIPT, the total energy of the ground state is 
\ba
E_1 = E_1^{(0)} + E_1^{(1)} + E_1^{(2)} + ...,
\ea
where the $k$'th order $E^{(k)}$ is proportional to $\mathcal{E}^k$, therefore the polarizability is 
the second order correction to the energy in TIPT 
\ba
\alpha = \frac{d^2E_1^{(2)}}{d\mathcal{E}^2} = \frac{d^2}{d\mathcal{E}^2}\sum_{n\ne 1} \frac{|\bra{\psi_1^{(0)}}V\ket{\psi_n^{(0)}}|^2}{E_1^{(0)}-E_n^{(0)}}.
\ea
Thus the system is more polarizable when the gaps between the ground and excited states are low.
Using Eqs.~\eqref{eq:Vn1} and \eqref{eq:gap} the expression for polarizability can be further simplified as
\ba
\alpha &= \frac{-64mL^4}{\pi^6}\sum_{n \neq 1}\frac{{n^2}[(-1)^{n}+1]^2}{(n^2-1)^5}
\ea
This infinite sum can be evaluated by the {\it Mathematica} program\cite{Mathematica}
\ba\label{eq:alpha}
\alpha &= 
\frac{(\pi^2 -15)}{12\pi^4}mL^4.
\ea
Therefore, the system polarizability grows linearly with the particle mass and quartically with the size of the box. 

\subsection{Charge mobility and oscillation amplitude}

More advanced but closely-related characterization of dynamics can be 
done using the conventional definition of the electron mobility $\mu_e=v_e/\mathcal{E}$,
where 
\ba
v_e = \frac{d}{dt}\bra{\psi(t)} x \ket{\psi(t)}, 
\ea
is the average electron velocity under the influence of the electric field $\mathcal{E}$.
Employing the Ehrenfest theorem\cite{Tannor:2007/35} it is easy to 
show that for our system
\ba
\frac{d}{dt}\bra{\psi(t)} x \ket{\psi(t)} &= \frac{\bra{\psi(t)} p \ket{\psi(t)}}{m} \\
\frac{d}{dt}\bra{\psi(t)} p \ket{\psi(t)} &= -\bra{\psi(t)}\frac{dV}{dx} \ket{\psi(t)} = \mathcal{E}.
\ea
The electron mobility is related to the electron conductivity ($\sigma$) with the simple relation 
$\sigma = ne\mu_e$, where $n$ is the number of electrons and $e$ their charge. 
The equations for the mobility and conductivity are appropriate for open systems where the steady 
flow of electrons is possible. In our closed system the dynamics will always have an oscillatory nature. 
To characterize its extent we will take the maximum value over the period 
\bea
\mu_e &=& \max_t\frac{d}{dt}\bra{\psi(t)} x \ket{\psi(t)}/{\mathcal{E}} \\
\label{eq:mudef}
&=& \frac{1}{m\mathcal{E}}\max_t \bra{\psi(t)} p \ket{\psi(t)}. 
\eea
In addition, since we deal with the oscillatory motion, the mobility can be characterized 
not only by velocity but also with the amplitude of the oscillation 
\bea\label{eq:Adef}
A =  \max_t |\bra{\psi(0)} x \ket{\psi(0)} -\bra{\psi(t)} x \ket{\psi(t)}|/L.
\eea
To understand what parameters determine $\mu_e$ and $A$ we will employ the 
time-dependent perturbation theory (TDPT) for the 
wave function time dependence 
\ba
\psi(x,t) = \psi^{(0)}(x,t) + \psi^{(1)}(x,t) + ...
\ea
Introducing the TDPT expansion for the wave function in Eqs.~\eqref{eq:mudef} and \eqref{eq:Adef}
and restricting consideration up to the 1st order we obtain the following equations
\ba
\mu_e^{(0+1)} &= \frac{1}{m\mathcal{E}}\max_t [\bra{\psi^{(0)}(t)} p \ket{\psi^{(1)}(t)} + ~h.c.]  \\
&=  \frac{2}{m\mathcal{E}}\max_t {\rm Re}[\bra{\psi^{(0)}(t)} p \ket{\psi^{(1)}(t)}] \\
A^{(0+1)} &=  \frac{1}{L}\max_t |\bra{\psi^{(0)}(t)} x \ket{\psi^{(1)}(t)} + ~ h.c.|. \\
&=  \frac{2}{L}\max_t {\rm Re}[\bra{\psi^{(0)}(t)} x \ket{\psi^{(1)}(t)}],
\ea
where $h.c.$ is the Hermitian conjugate, and the following easily verifiable 
relations are used
\ba
\bra{\psi^{(0)}(t)} p \ket{\psi^{(0)}(t)} &= 0, \\
\bra{\psi(0)} x \ket{\psi(0)} &=  \bra{\psi^{(0)}(t)} x \ket{\psi^{(0)}(t)}.
\ea
Following the 1st order TDPT, the $\psi^{(1)}(x,t)$ will be given by 
\ba
\psi^{(1)}(x,t)  &= \sum_{n\ne 1} d_n^{(1)}(t) \psi^{(0)}_n(x)e^{-iE_n^{(0)}t} \\
d_n^{(1)}(t) &= i\mathcal{E} \int_0^t \bra{\psi^{(0)}_1} x \ket{\psi^{(0)}_n} e^{i(E_n^{(0)}-E_1^{(0)})\tau} d\tau \\
&= \frac{\mathcal{E}\bra{\psi^{(0)}_1} x \ket{\psi^{(0)}_n}}{E_n^{(0)}-E_1^{(0)}} (e^{i(E_n^{(0)}-E_1^{(0)})t}-1)
\ea 
Combining all the terms gives 
\bea\notag
\psi^{(1)}(x,t) &=& \sum_{n\ne 1} \psi^{(0)}_n(x) \frac{\mathcal{E}\bra{\psi^{(0)}_1} x \ket{\psi^{(0)}_n}}{E_n^{(0)}-E_1^{(0)}} \\
&&\times(e^{-iE_1^{(0)}t}-e^{-iE_n^{(0)}t})
\eea
Using this in combination with $\psi^{(0)}(x,t) = e^{-iE_1^{(0)}t}\psi_1^{(0)}(x)$ allows us to 
express the mobility as a series
\bea\label{eq:mu}
\mu_e^{(0+1)} &=& \frac{32L^2}{\pi^4}\max_t\bigg\{\sum_{n \neq 1}^{\infty}\frac{n^2[(-1)^n+1]^2}{(n^2-1)^4}\\ \notag
&&\times\sin\left[\frac{t\pi^2(n^2-1)}{2mL^2}\right]\bigg\}
\eea
As the degree of $n$ is much higher in the denominator than the numerator, the $n=2$ term will have the largest effect. If we take only this term, we can maximize the sine function by choosing a value of $t$ such that
\ba{\label{eq:mu_t}}
t = \frac{mL^2(4k+1)}{3\pi}, k \in \mathbb{Z}.
\ea
Using these values in the $n=2$ term to obtain an approximate value for the maximum charge mobility,
\ba{\label{eq:mu_approx}}
\mu_e^{(0+1)} \approx \frac{512L^2}{81\pi^4}.
\ea
This expression is independent of mass, showing that while the charge mobility varies quadratically 
with the size of the box, it is not affected by the mass of the particle. 
This independence from the mass is somewhat counterintuitive from the classical point of 
view because one would expect heavier particles to be slower 
and lighter particles to be faster. 
However, in the PIB model, gaps between the ground and excited states are 
inversely proportional to the mass [\eq{eq:gap}], 
which generally makes the mobility higher as mass increases. On the other hand, the mass dependence 
that is in \eq{eq:mudef} provides the opposite trend. The two trends cancel each other,
removing the mass dependence from the mobility.  

Similar to the mobility, the oscillation amplitude can be expressed as a series
\bea\label{eq:A}
A^{(0+1)}  &=&  \frac{64\mathcal{E}mL^3}{\pi^6}\max_t\bigg[\sum_{n \neq 1}^{\infty}\frac{n^2[(-1)^n + 1]^2}{(n^2-1)^5}\\ \notag
&&\times\left(1-\cos\left(\frac{t\pi^2(n^2-1)}{2mL^2}\right)\right)\bigg].
\eea
Again, the terms decrease in magnitude as $n$ increases, so the $n=2$ term will have the greatest contribution to the overall sum. For this term, the ideal value of $t$ can be expressed as
\ba{\label{eq:A_t}}
t = \frac{2mL^2(2k+1)}{3\pi}, k \in \mathbb{Z}
\ea
With these values in the $n=2$ term, an approximate value for the amplitude is
\ba{\label{eq:A_approx}}
A^{(0+1)} \approx \frac{2048\mathcal{E}mL^3}{243\pi^6}.
\ea
In contrast with the mobility, the amplitude not only grows with $L$ but also increases linearly with $m$ and $\mathcal{E}$. 

\section{Numerical Illustration}

All quantities that we have considered in connection with the electron transport in the PIB system
are inversely proportional to the gap between the ground and excited states. 
This gap is determined by two parameters: $m$ and $L$ [\eq{eq:gap}]. 
Here we illustrate the dependence of the considered properties on these two parameters using 
the variational approach to obtain eigenstates of the PIB problem with a finite field. 
Results of variational calculations will be compared to perturbative estimates obtained in the Theory section.

\begin{figure}
\includegraphics[scale=0.45]{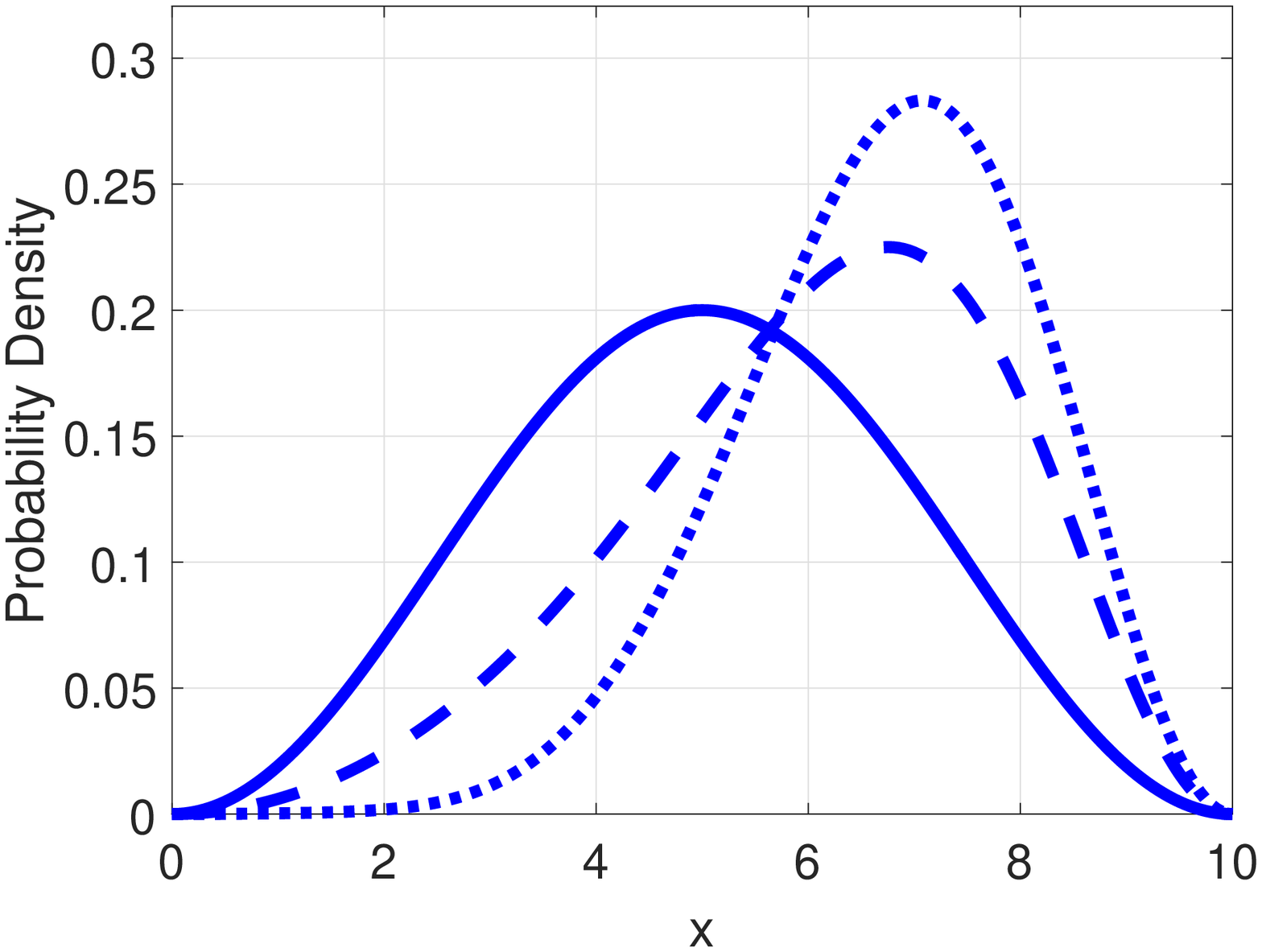}
\includegraphics[scale=0.45]{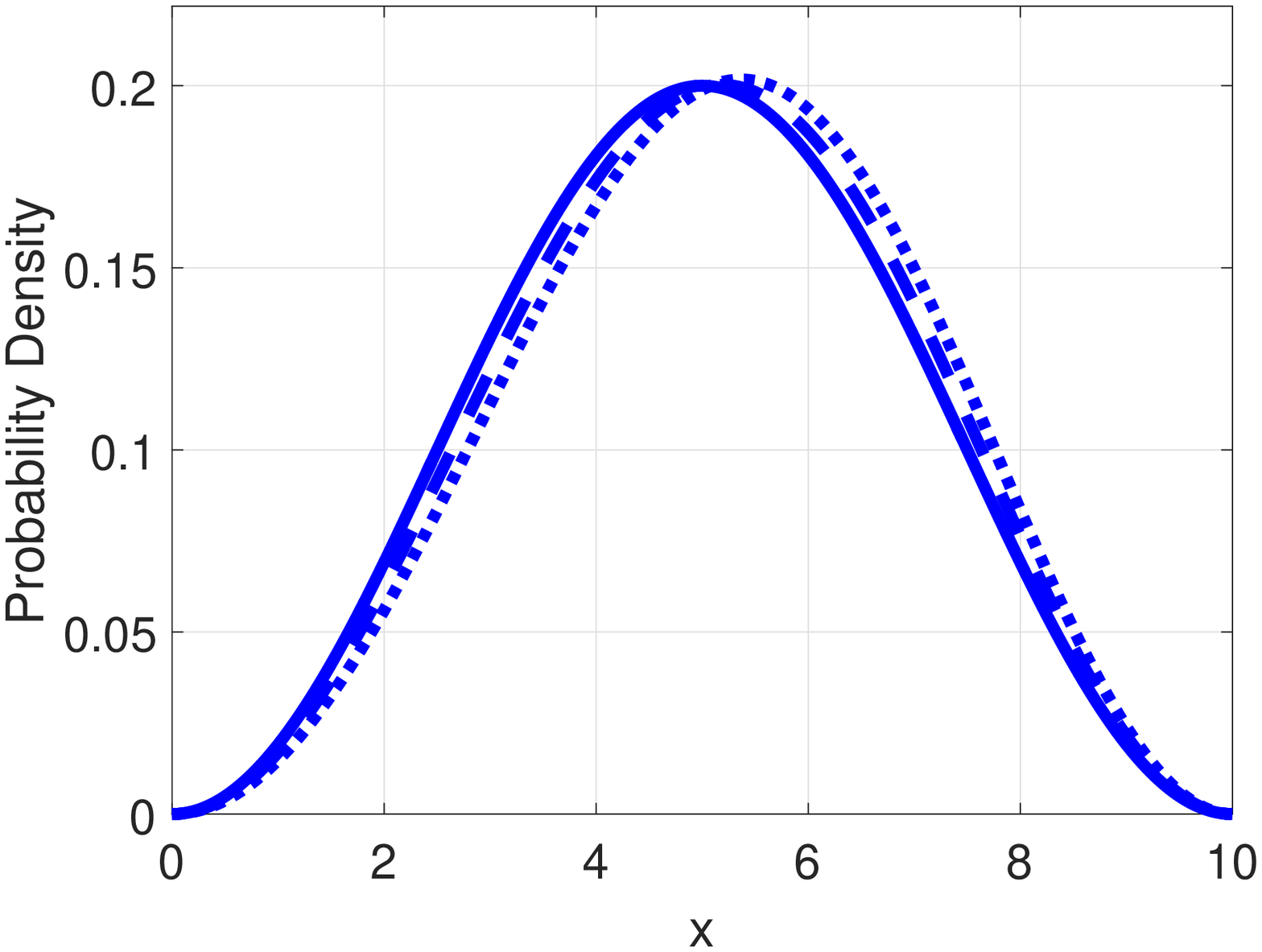}
\caption{Probability density as a function of the $x$-coordinate for $m = 0.5$ (upper plot), $m = 0.05$ (lower plot)
at different times:  $t = 0$ (solid), $t = t_{\rm max}/2$ (dashed), $t = t_{\rm max}$ (dotted). $L = 10$, $\mathcal{E} = 0.05$, 
$t_{\rm max}\approx 200m/(3\pi)$, and 20 eigenfunctions were used for both systems.}\label{fig:osc}
\end{figure}

\paragraph{Probability density oscillations:} 
One of the simplest demonstrations of the relation between conductivity 
and the gap is to observe the probability density dynamics for two 
particles of different masses (see \fig{fig:osc}). We build these
dynamics using \eq{eq:dist} to evaluate $|\psi(x,t)|^2$ and plot it as a function of $x$ 
for a few times. The lower mass system has a larger gap (see \eq{eq:gap}) 
and oscillations of its probability density are diminished 
compare to those in the heavy mass system, where the gap is much smaller.

\begin{figure}
\includegraphics[scale=0.45]{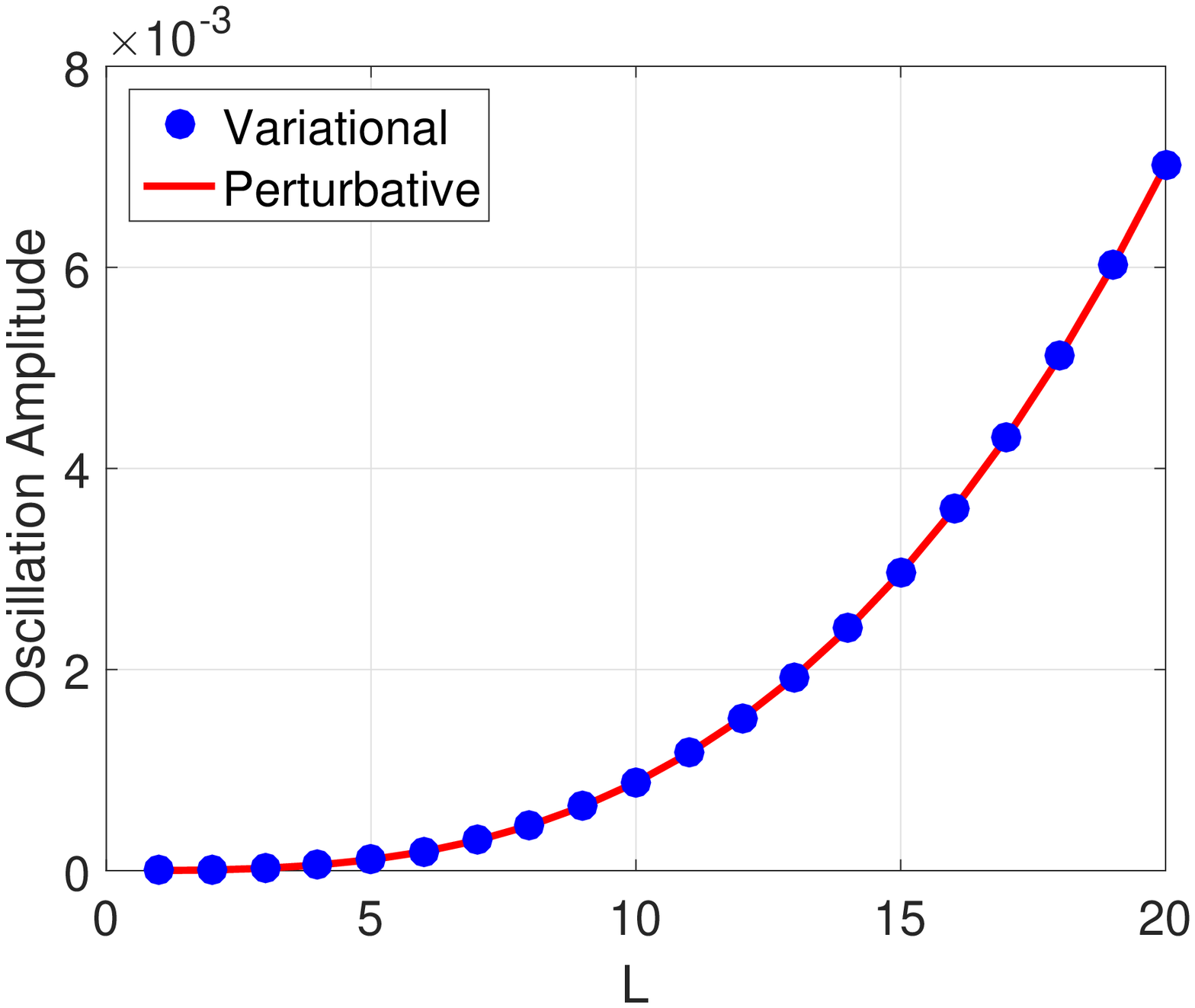}
\includegraphics[scale=0.45]{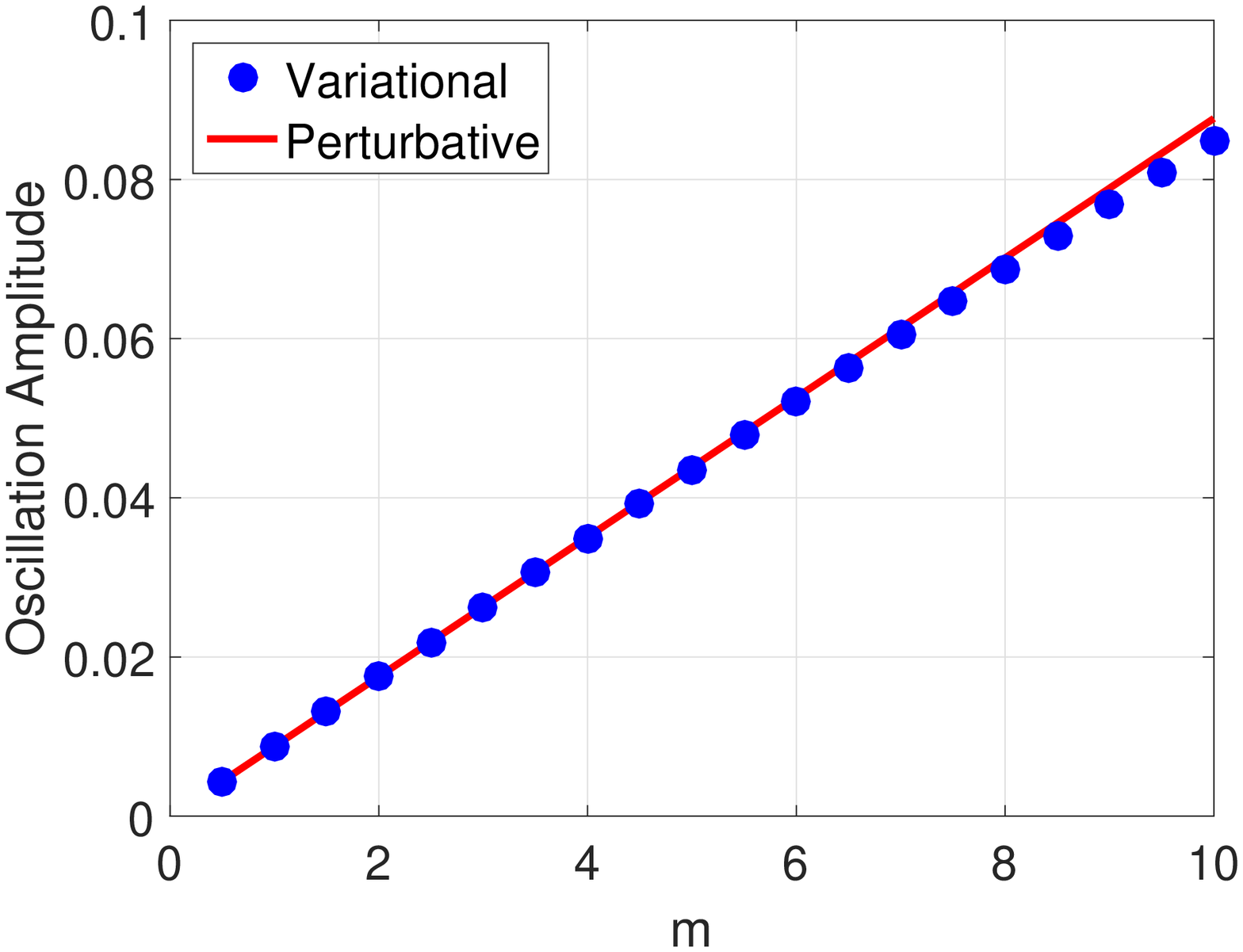}
\includegraphics[scale=0.45]{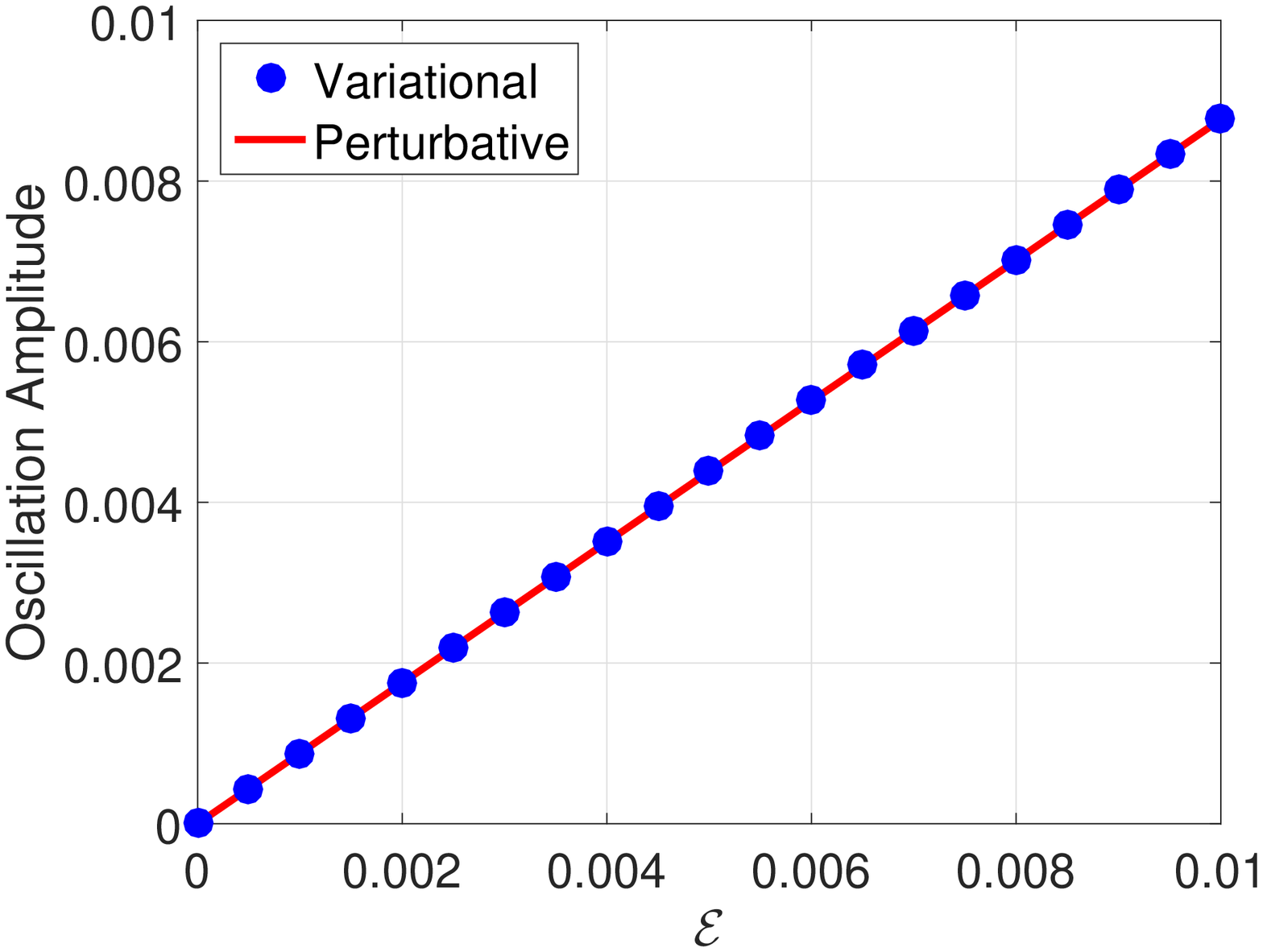}
\caption{Maximum amplitude as a function of $L$ (upper plot), 
$m$ (middle plot), and $\mathcal{E}$ (lower plot). Fixed values were 
$m = 0.1$, $\mathcal{E} = 0.001$, and $L = 10$.}\label{fig:amp}
\end{figure}

According to \eq{eq:A_approx}, the amplitude of oscillations 
is dependent on $\mathcal{E}$, $m$, and $L$, and \fig{fig:amp}
summarizes our comparison of the amplitude obtained using variational and perturbative expressions.
 As evident from \fig{fig:amp}, the variational method's 
results agree with \eq{eq:A_approx} from the perturbative approach, assuming 
the perturbation is not large. Both approaches reveal that amplitude increases with $L$,
since increasing $L$ decreases the gap and makes system more susceptible to oscillations. 
Note that this does not simply happen because the
 larger box gives the particle more room to move; such logic would be incorrect because
 we consider the amplitude normalized by the size of the box [\eq{eq:Adef}]. One can also see an 
obvious trend of increased amplitudes for heavier masses, which again stems from the 
reduction of the gap for increased mass, described by \eq{eq:gap}. Finally, there is a 
dependence on $\mathcal{E}$, as a larger applied potential causes a greater 
oscillation amplitude.

\begin{figure}
\includegraphics[scale=0.45]{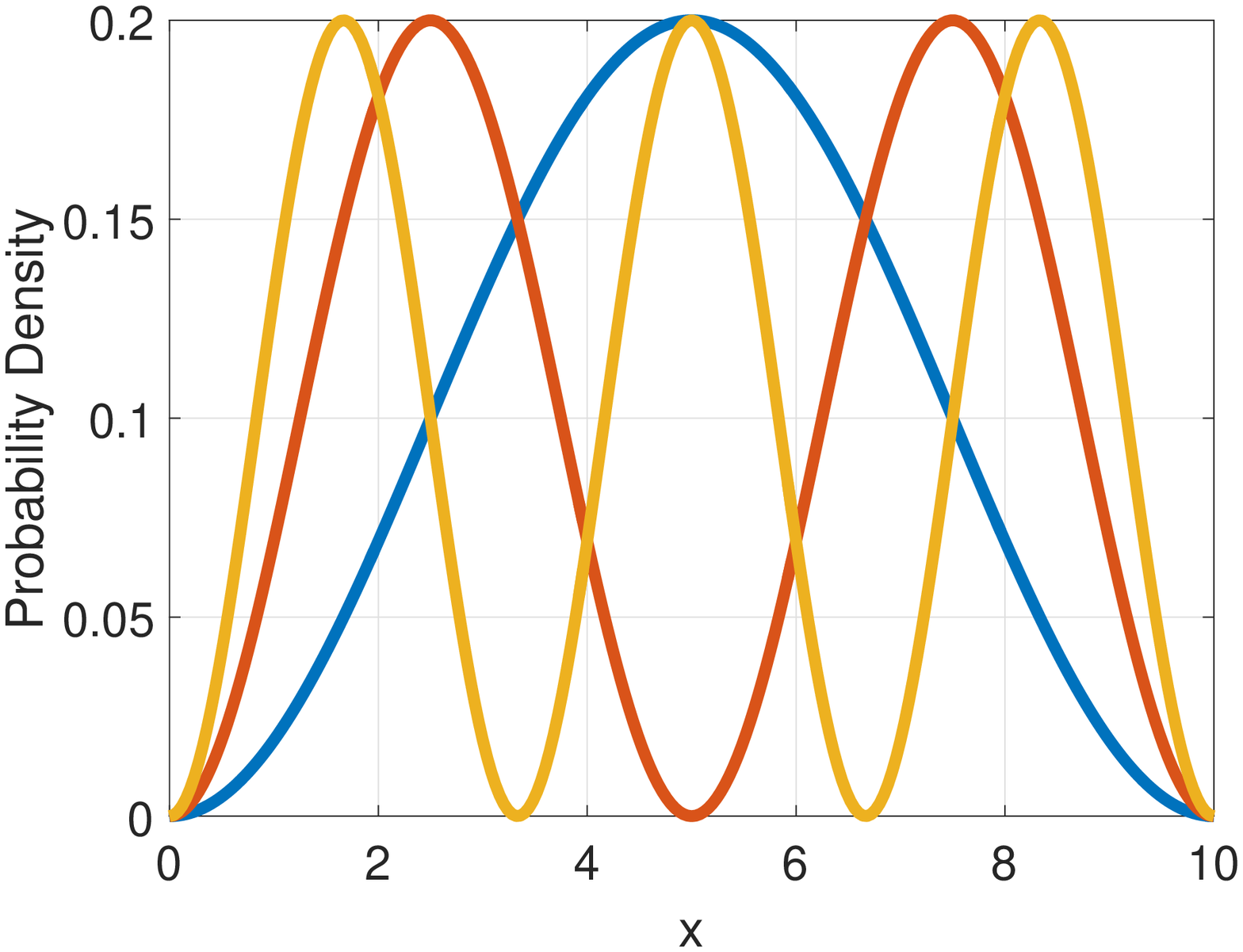}
\includegraphics[scale=0.45]{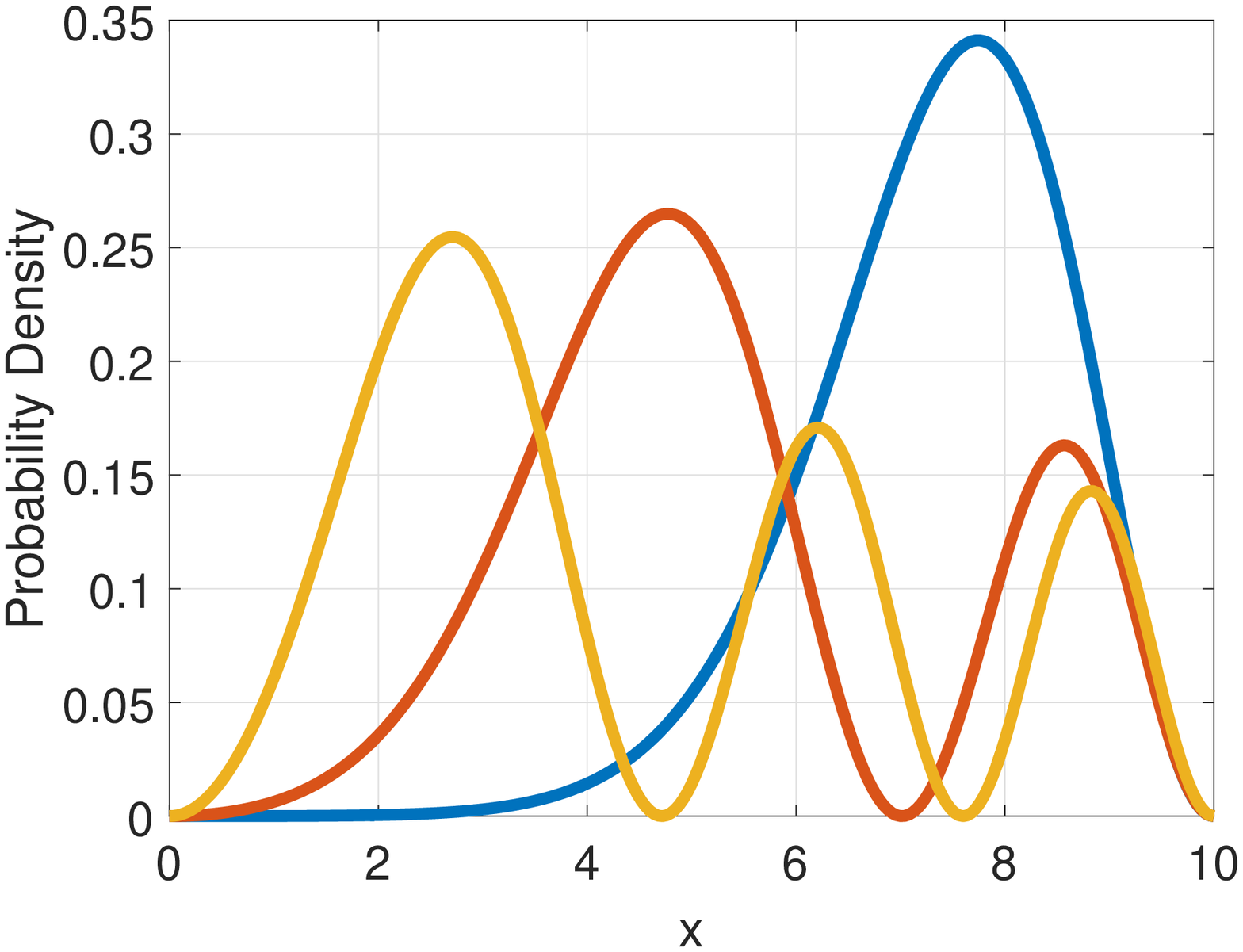}
\caption{The probability densities for the first three eigenfunctions of PIB problems, 
where $m = 1$ and $L = 10$: ground (blue), first (red) and second (green) excited states. 
The top (bottom) plot corresponds to eigenstates of $H_0$ ($H$ with $\mathcal{E}=0.1$).}\label{fig:eigs}
\end{figure}

\paragraph{Eigenstates for the PIB model with an inclined bottom:}
To understand quantum interference of what states leads to the oscillations 
depicted in \fig{fig:osc} we consider the eigenstates of $H$ in \eq{eq:H}. 
Note that the key role in spatial extent of the oscillations is the spatial distribution 
of probability density for individual states,
because if all states forming a superposition are localized in a certain region of space, 
the interference dynamics cannot leave that region.
Figure \ref{fig:eigs} compares the three lowest eigenstates for the PIB model with and without inclination of the bottom. 
It is clear that inclination shifts the maximum of the probability distribution for 
the ground state closer to the region of lower potential. Interestingly, maxima of the probability density for 
the excited states shift the other way. This can be understood considering that all eigenstates must be mutually 
orthogonal. Therefore, if excited states followed the ground state they would need to have an increasing number of nodes
in the region of low potential to be orthogonal to the ground state, which would increase their kinetic energy.
Instead, they shift the other way in order to avoid introducing a dramatic curvature.

\begin{figure}
\includegraphics[scale=0.45]{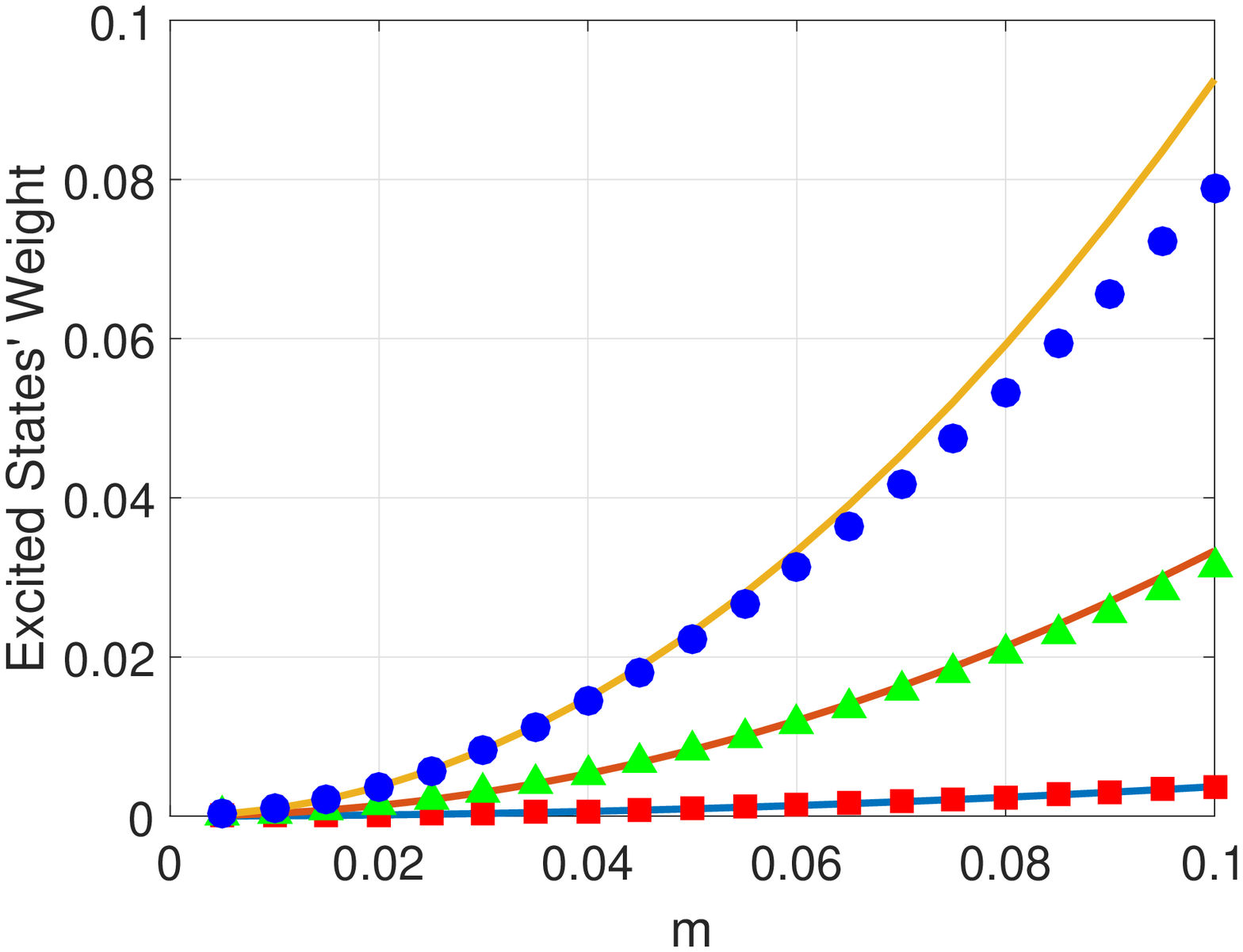}
\includegraphics[scale=0.45]{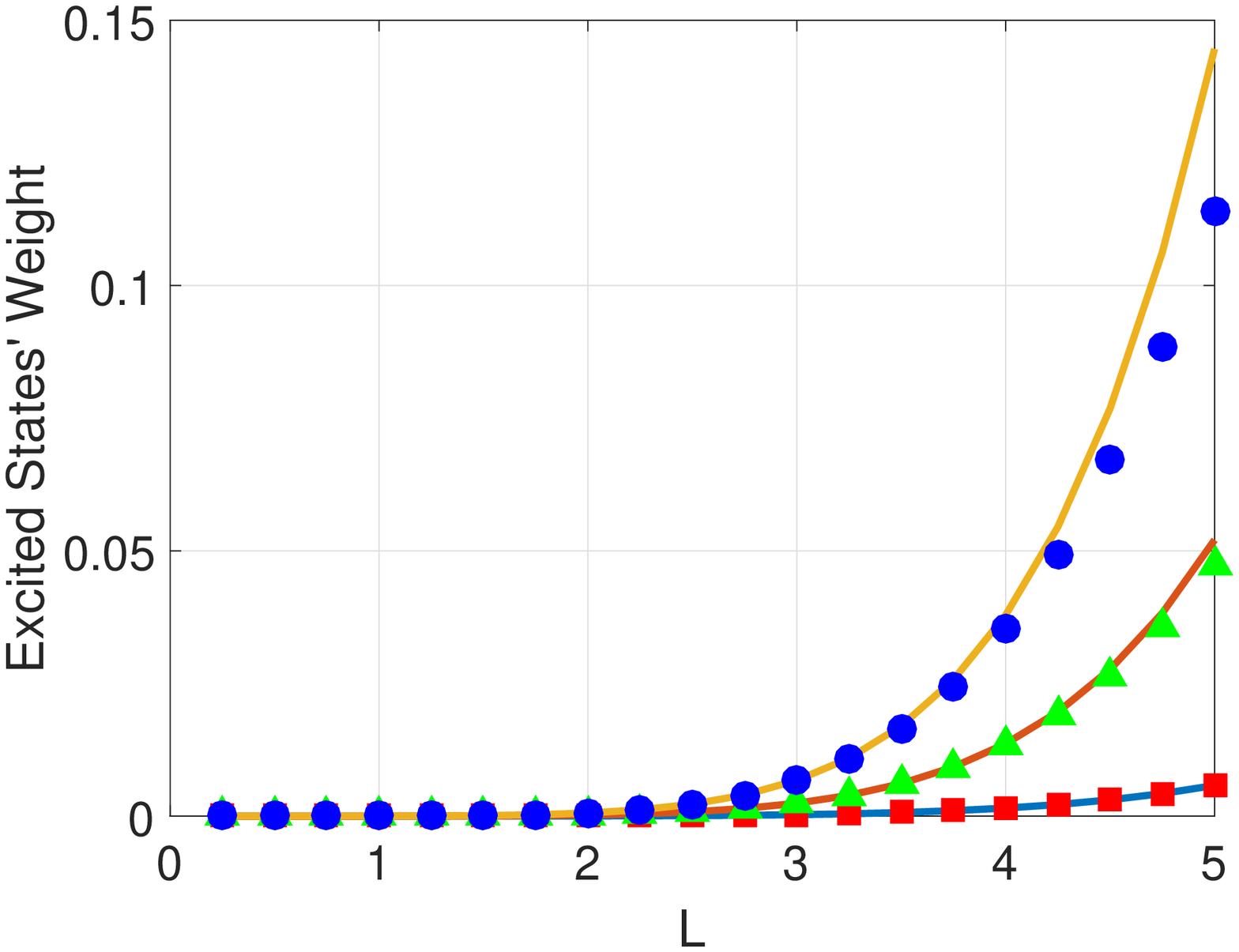}
\caption{The weight of excited states in the initial state ($\omega_{\rm exc}$) as a function of the particle 
mass (top panel, $L=10$) and as a function of the box size (bottom panel, $m=1$). Dots (solid lines) correspond to 
variational (perturbative [\eq{eq:omega}]) results for different field strengths: $\mathcal{E}=0.05$ squares, $\mathcal{E}=0.1$ triangles, and $\mathcal{E}=0.15$ circles.}\label{fig:esw}
\end{figure}

\paragraph{Weight of excited states:}
Another important component of the oscillatory dynamics is how many excited states are substantially 
involved in the superposition. Examining these states for the PIB model with the inclined bottom 
showed that they are generally shifted in the opposite side of the box compare to the shift of the 
ground state. However, if we do not have substantial weights of these states in the superposition, 
their localization will not be able to play a role in dynamics.
To characterize how the total weight of excited states 
changes within the initial state $\psi(x,0)=\psi_1^{(0)}(x)$ we evaluate 
$\omega_{\rm exc} = \sum_{n=2}^{20} |c_n|^2$ dependence on $L$ and $m$ using the variational 
approach [see \fig{fig:esw}]. Going beyond 20 states in the $\omega_{\rm exc}$ expansion 
does not produce non-negligible contribution in the range of reported parameters.
The perturbative estimate of $\omega_{\rm exc}$ in \eq{eq:omega} is in very good agreement 
with variational results. As the electric field strength increases, the first order perturbative estimate in \eq{eq:omega} 
becomes less accurate [\fig{fig:esw}] but maintains qualitatively correct trends. 
The origin of growth of $\omega_{\rm exc}$ with $L$ and $m$ is the reducing gap between 
the ground and excited states.

\begin{figure}
\includegraphics[scale=0.45]{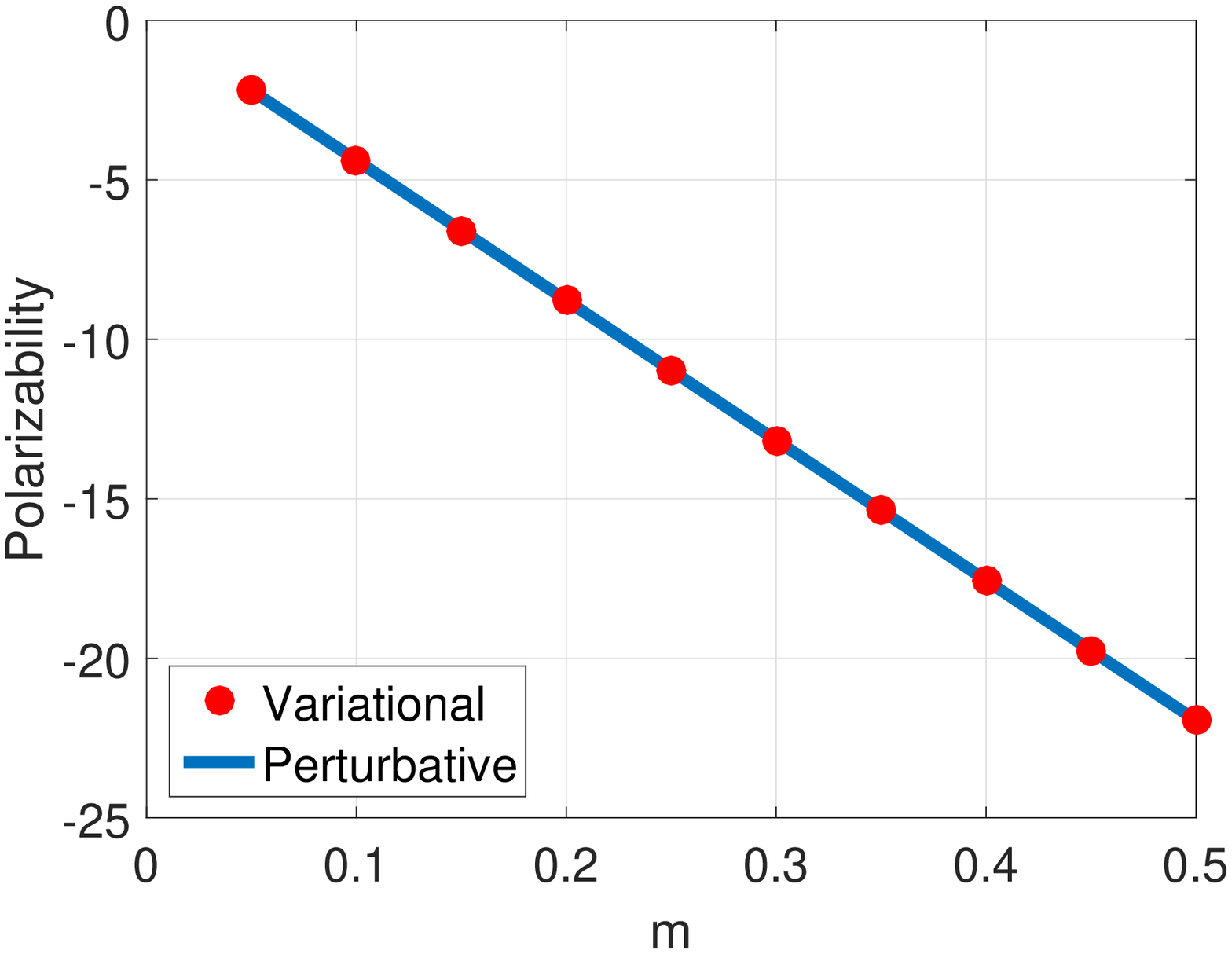}
\includegraphics[scale=0.45]{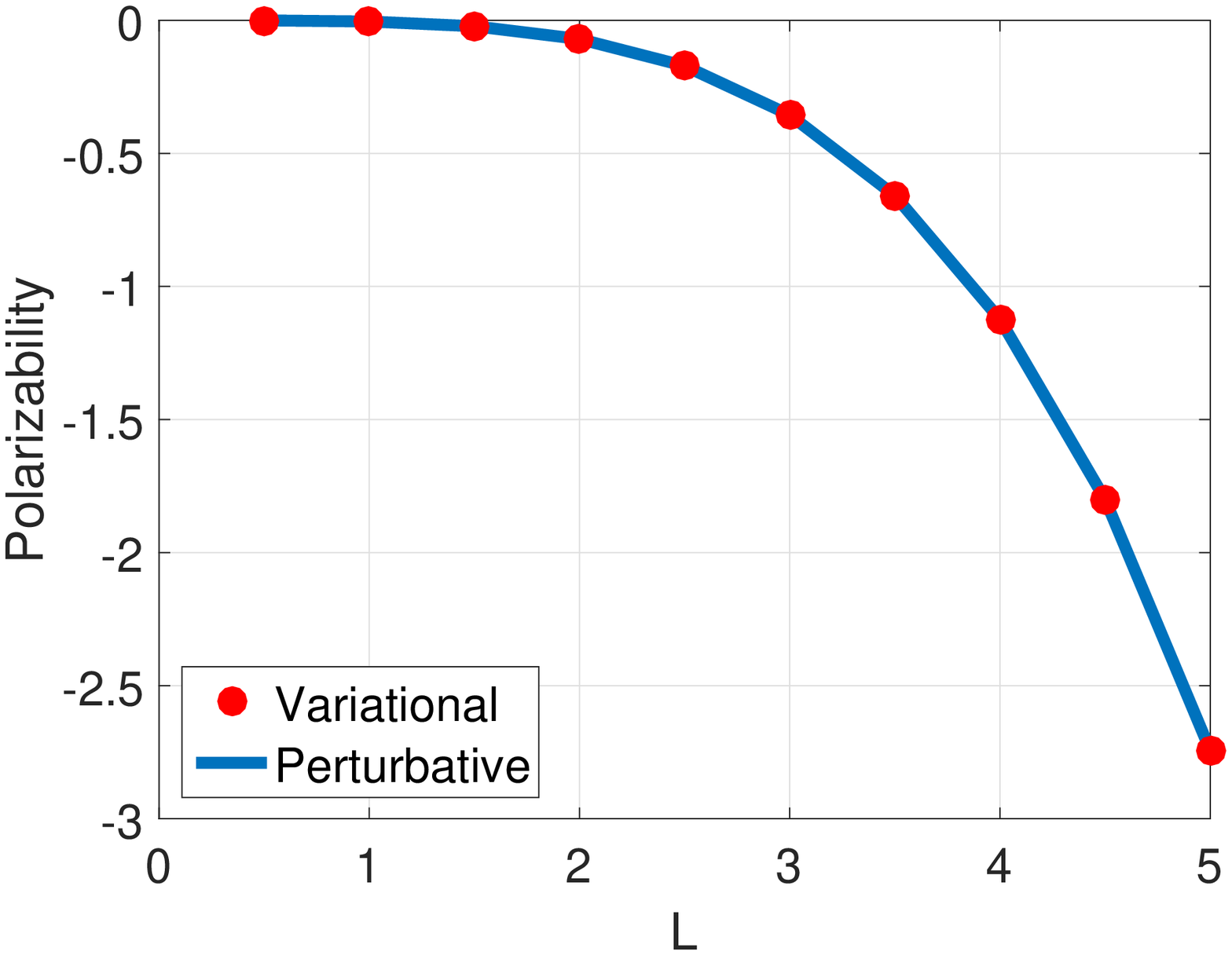}
\caption{Polarizability as a function of the particle 
mass (top panel, $L=10$) and as a function of the box size (bottom panel, $m=1$). 
Dots (solid lines) correspond to the variational (perturbative [\eq{eq:alpha}]) approach.}\label{fig:pol}
\end{figure}

\paragraph{Polarizability:} 
This quantity characterizes the response of the system to an infinitesimal electric field. 
The perturbation theory gives its exact value and to illustrate this numerically we 
compared perturbative expressions [\eq{eq:alpha}] with estimates 
obtained via numerical second derivatives of variational energies.
The energy differentiation was accomplished via the central finite 
differencing scheme within the eighth-order expression\cite{fornberg:1988} 
\bea\notag
\alpha &\approx& \bigg(\frac{-1}{560}E(-4\mathcal{E})+
\frac{8}{315}E(-3\mathcal{E})-\frac{1}{5}E(-2\mathcal{E})+\frac{8}{5}E(-\mathcal{E})\\ \notag
&&-\frac{205}{72}E(0)+\frac{8}{5}E(\mathcal{E})-\frac{1}{5}E(2\mathcal{E})+\frac{8}{315}E(3\mathcal{E})\\
&&-\frac{1}{560}E(4\mathcal{E})\bigg)/\mathcal{E}^2+O(\mathcal{E}^{8}),
\eea
where $E(\mathcal{E})$ is the ground state variational energy 
obtained at the $\mathcal{E}$ field value, $\mathcal{E}=10^{-4}$
a.u. was used. 
Figure \ref{fig:pol} shows that perturbative and variational polarizability results are seemingly indistinguishable,
both approaches display linear dependence on $m$ and quartic dependence on $L$.

\begin{figure}
\includegraphics[scale=0.45]{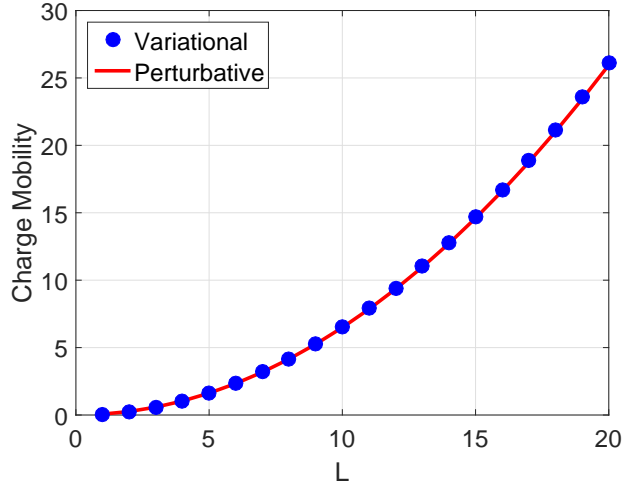}
\caption{Charge mobility dependence on the length of the box, 
for $\mathcal{E} = 0.001$ and $m = 0.1$. 
The variational results were obtained using the first 15 eigenfunctions.}\label{fig:mob}
\end{figure}

\paragraph{Charge mobility:}
According to \eq{eq:mu_approx} the charge mobility is independent of mass and 
can only be changed by varying the size of the box $L$. 
To model mobility in \fig{fig:mob}, we evaluate $\mu_e$ using \eq{eq:mudef} 
for different values of $L$, $m$, and $\mathcal{E}$. 
To determine the average momentum, ${\psi}(x,t)$ was calculated using the variational approach. 
To ensure that we picked times of high momentum, we use $t$ from \eq{eq:mu_t} 
with $k = 1$, maximizing \eq{eq:mu} where $n=2$. 
The resulting discrete values were plotted along with the solutions to \eq{eq:mu_approx}, 
and their clear similarity shows that as long as the perturbation is sufficiently small, 
mobility is quadratically dependent on $L$. Since the plots for $m$ and $\mathcal{E}$ were flat lines, 
they were not included in this consideration, though these results agree with the perturbative 
prediction that mobility does not vary with mass or field strength.

\section{Concluding Remarks}
For a quantum particle to change its probability density in time, the particle needs to be 
in a superposition state. This is one of the most important concepts that underlies the 
phenomenon of conductivity. The particle-in-a-box problem with an inclined bottom 
can be used to illustrate this basic quantum mechanical principle behind electric conductivity. 
The key element in this phenomenon is that electrons in conductive 
materials are in the non-stationary state after the application of an electric field.  
The crucial system parameter for an efficient creation of this non-stationary (superposition) state is 
the gap between ground and excited electronic states. The larger the gap, 
the more energy required to create the superposition state necessary for the particle transport.
 
The particle-in-a-box model is the simplest model system where this dependence can be illustrated. 
This model is a useful example for lecture demonstrations using computer programs, for example MatLab.\cite{Matlab:2012b} 
It can also serve as a source of computer-assisted problems for computer-literate undergraduate 
students to obtain hands-on experience with the variational approach and perturbation theory.

\section*{Acknowledgements}

A.F.I. is grateful to Al-Amin Dhirani, Dvira Segal, and Mark Ratner for stimulating discussions
and acknowledges funding from the Natural Sciences and Engineering 
Research Council of Canada (NSERC) through the Discovery Grants Program 
and the Alfred P. Sloan Foundation. 


\begin{thebibliography}{9}%
\makeatletter
\providecommand \@ifxundefined [1]{%
 \@ifx{#1\undefined}
}%
\providecommand \@ifnum [1]{%
 \ifnum #1\expandafter \@firstoftwo
 \else \expandafter \@secondoftwo
 \fi
}%
\providecommand \@ifx [1]{%
 \ifx #1\expandafter \@firstoftwo
 \else \expandafter \@secondoftwo
 \fi
}%
\providecommand \natexlab [1]{#1}%
\providecommand \enquote  [1]{``#1''}%
\providecommand \bibnamefont  [1]{#1}%
\providecommand \bibfnamefont [1]{#1}%
\providecommand \citenamefont [1]{#1}%
\providecommand \href@noop [0]{\@secondoftwo}%
\providecommand \href [0]{\begingroup \@sanitize@url \@href}%
\providecommand \@href[1]{\@@startlink{#1}\@@href}%
\providecommand \@@href[1]{\endgroup#1\@@endlink}%
\providecommand \@sanitize@url [0]{\catcode `\\12\catcode `\$12\catcode
  `\&12\catcode `\#12\catcode `\^12\catcode `\_12\catcode `\%12\relax}%
\providecommand \@@startlink[1]{}%
\providecommand \@@endlink[0]{}%
\providecommand \url  [0]{\begingroup\@sanitize@url \@url }%
\providecommand \@url [1]{\endgroup\@href {#1}{\urlprefix }}%
\providecommand \urlprefix  [0]{URL }%
\providecommand \Eprint [0]{\href }%
\providecommand \doibase [0]{http://dx.doi.org/}%
\providecommand \selectlanguage [0]{\@gobble}%
\providecommand \bibinfo  [0]{\@secondoftwo}%
\providecommand \bibfield  [0]{\@secondoftwo}%
\providecommand \translation [1]{[#1]}%
\providecommand \BibitemOpen [0]{}%
\providecommand \bibitemStop [0]{}%
\providecommand \bibitemNoStop [0]{.\EOS\space}%
\providecommand \EOS [0]{\spacefactor3000\relax}%
\providecommand \BibitemShut  [1]{\csname bibitem#1\endcsname}%
\let\auto@bib@innerbib\@empty
\bibitem [{\citenamefont {Fayer}(2010)}]{Fayer:2013/73}%
  \BibitemOpen
  \bibfield  {author} {\bibinfo {author} {\bibfnamefont {M.~D.}\ \bibnamefont
  {Fayer}},\ }in\ \href@noop {} {\emph {\bibinfo {booktitle} {Absolutely Small:
  How Quantum Theory Explains Our Everyday World}}}\ (\bibinfo  {publisher}
  {Amacom},\ \bibinfo {address} {New York},\ \bibinfo {year} {2010})\ p.\
  \bibinfo {pages} {329}\BibitemShut {NoStop}%
\bibitem [{\citenamefont {Engel}(2013)}]{Engel:2013/73}%
  \BibitemOpen
  \bibfield  {author} {\bibinfo {author} {\bibfnamefont {T.}~\bibnamefont
  {Engel}},\ }in\ \href@noop {} {\emph {\bibinfo {booktitle} {Quantum Chemistry
  and Spectroscopy}}}\ (\bibinfo  {publisher} {Pearson},\ \bibinfo {address}
  {Boston},\ \bibinfo {year} {2013})\ p.~\bibinfo {pages} {73}\BibitemShut
  {NoStop}%
\bibitem [{\citenamefont {Ziman}(1960)}]{Ziman:1960/92}%
  \BibitemOpen
  \bibfield  {author} {\bibinfo {author} {\bibfnamefont {J.~M.}\ \bibnamefont
  {Ziman}},\ }in\ \href@noop {} {\emph {\bibinfo {booktitle} {Electrons and
  Phonons: The Theory of Transport Phenomena in Solids}}}\ (\bibinfo
  {publisher} {Oxford at the Clarendon Press},\ \bibinfo {address} {Oxford},\
  \bibinfo {year} {1960})\ p.~\bibinfo {pages} {92}\BibitemShut {NoStop}%
\bibitem [{\citenamefont {Ashcroft}\ and\ \citenamefont
  {Mermin}(1976)}]{Ashcroft:1976/214}%
  \BibitemOpen
  \bibfield  {author} {\bibinfo {author} {\bibfnamefont {N.~W.}\ \bibnamefont
  {Ashcroft}}\ and\ \bibinfo {author} {\bibfnamefont {N.~D.}\ \bibnamefont
  {Mermin}},\ }in\ \href@noop {} {\emph {\bibinfo {booktitle} {Solid State
  Physics}}}\ (\bibinfo  {publisher} {Saunders College},\ \bibinfo {address}
  {Orlando, Florida},\ \bibinfo {year} {1976})\ p.\ \bibinfo {pages}
  {214}\BibitemShut {NoStop}%
\bibitem [{\citenamefont {Wolfram~Research}(2016)}]{Mathematica}%
  \BibitemOpen
  \bibfield  {author} {\bibinfo {author} {\bibfnamefont {I.}~\bibnamefont
  {Wolfram~Research}},\ }\href@noop {} {\enquote {\bibinfo {title}
  {Mathematica, version 10.4},}\ }\bibinfo {howpublished} {Wolfram Research,
  Inc., Champaign, Illinois} (\bibinfo {year} {2016})\BibitemShut {NoStop}%
\bibitem [{\citenamefont {Kutzelnigg}(2006)}]{Kutzelnigg:2006ij}%
  \BibitemOpen
  \bibfield  {author} {\bibinfo {author} {\bibfnamefont {W.}~\bibnamefont
  {Kutzelnigg}},\ }\href@noop {} {\bibfield  {journal} {\bibinfo  {journal}
  {Journal of Computational Chemistry}\ }\textbf {\bibinfo {volume} {28}},\
  \bibinfo {pages} {25} (\bibinfo {year} {2006})}\BibitemShut {NoStop}%
\bibitem [{\citenamefont {Tannor}(2007)}]{Tannor:2007/35}%
  \BibitemOpen
  \bibfield  {author} {\bibinfo {author} {\bibfnamefont {D.~J.}\ \bibnamefont
  {Tannor}},\ }in\ \href@noop {} {\emph {\bibinfo {booktitle} {Introduction to
  Quantum Mechanics: A Time-Dependent Perspective}}}\ (\bibinfo  {publisher}
  {University Science Books},\ \bibinfo {address} {Sausalito, California},\
  \bibinfo {year} {2007})\ p.~\bibinfo {pages} {35}\BibitemShut {NoStop}%
\bibitem [{\citenamefont {Fornberg}(1988)}]{fornberg:1988}%
  \BibitemOpen
  \bibfield  {author} {\bibinfo {author} {\bibfnamefont {B.}~\bibnamefont
  {Fornberg}},\ }\href@noop {} {\bibfield  {journal} {\bibinfo  {journal}
  {Mathematics of Computation}\ }\textbf {\bibinfo {volume} {51}},\ \bibinfo
  {pages} {699} (\bibinfo {year} {1988})}\BibitemShut {NoStop}%
\bibitem [{Mat(2012)}]{Matlab:2012b}%
  \BibitemOpen
  \href@noop {} {\enquote {\bibinfo {title} {Matlab, version 8.0.0.783
  (r2012b)},}\ }\bibinfo {howpublished} {The MathWorks Inc., Natick,
  Massachusetts} (\bibinfo {year} {2012})\BibitemShut {NoStop}%
\end{thebibliography}
%

\end{document}